\documentclass[aps,prd,reprint,preprintnumbers,superscriptaddress,showpacs,twocolumn]{revtex4-1}
\usepackage{latexsym,graphicx,amssymb,amsmath,mathrsfs}
\usepackage{setspace,bm}
\usepackage[breaklinks, colorlinks=true, pdfstartview=FitV, linkcolor=red, citecolor=blue, urlcolor=blue]{hyperref}
\usepackage[usenames]{color}
\usepackage{epstopdf}
\usepackage{mathtools}
\usepackage{comment}
\usepackage{braket}
\usepackage{CJKutf8}
\usepackage{physics}
\usepackage{slashed}
\usepackage{bbold}
\usepackage{tikz}
\usetikzlibrary{shapes.geometric,arrows,calc}
\usetikzlibrary{arrows.meta}

\begin{document}

\title{Information geometry of non-equilibrium quantum states:\\
       Mixed metric structures on an extended information manifold}

\author{Kouji Kashiwa}
\email[]{kashiwa@fit.ac.jp}
\affiliation{Department of Computer Science and Engineering, Faculty of Information Engineering, Fukuoka Institute of Technology, Fukuoka 811-0295, Japan}

\author{Hidefumi Matsuda}
\email{da.matsu.00.bbb.kobe@gmail.com}
\affiliation{Zhejiang Institute of Modern Physics, Department of Physics, Zhejiang University, Hangzhou, 310027, China}

\begin{abstract}
We discuss an information-geometric framework for characterizing quantum states in non-equilibrium dynamics.
Using the transverse-field Ising chain model as a laboratory, we investigate the quantum Fisher information metric with particular emphasis on the mixed metric component $g_{ht}$ and related geometric observables, where $h$ is a controllable post-quench Hamiltonian parameter and $t$ is real time. The framework treats $h$ and $t$ as coordinates of an extended information manifold $(h,t)$. The geometric observables characterize the speed of evolution on the manifold and the alignment between the temporal and parameter-deformation directions. Correlations between these geometric quantities and two widely used measures of non-equilibrium dynamics, the entanglement entropy and the Loschmidt echo, are analyzed.
\end{abstract}

\maketitle

\section{Introduction}
\label{sec:introduction}
Understanding non-equilibrium quantum many-body systems is one of the central challenges in modern physics.
Thermalization~\cite{DAlessio:2015qtq}, pre-thermalization~\cite{Mori:2017qhg}, entanglement spreading~\cite{Eisert:2008ur}, and dynamical quantum phase transitions~\cite{Heyl:2017blm} provide rich examples of non-equilibrium phenomena arising in real-time quantum evolution.
In parallel, there has been growing interest in geometric descriptions of quantum many-body dynamics~\cite{Kolodrubetz:2017ofs}.
These ideas may be relevant not only to condensed-matter physics, but also to elementary particle physics.

Information-theoretic approaches have become powerful probes of quantum phases and phase transitions.
Among them, the quantum Fisher information (QFI) is important because it defines a natural Riemannian metric on the manifold of quantum states~\cite{Braunstein:1994zz}; see Ref.~\cite{Liu:2019xfr} for a review of the QFI.
In equilibrium systems, the QFI metric is closely related to fidelity susceptibility~\cite{You:2007ofl,Gu:2008dyh,Liu:2014mip,Liu:2019xfr} and often exhibits characteristic enhancement near quantum critical points; see Refs.~\cite{Zanardi:2006mgo,Venuti:2007izd,Zanardi:2007byf,Gu:2008dyh} as examples.
More generally, quantum geometric quantities play important roles in both adiabatic and non-adiabatic dynamics; see Ref.~\cite{Kolodrubetz:2017ofs} as an example.

Recent studies have begun to explore the role of the QFI in non-equilibrium quantum dynamics; see Ref.~\cite{guan2026exploring} as an example.
In particular, the post-quench evolution of the QFI in one-dimensional periodic systems has been investigated, where both diagonal and off-diagonal geometric components were related to physical quantities such as energy fluctuations, group-velocity variance, and quench-induced geometric structures~\cite{Tang:2026xap}.
In addition, a time-dependent extension of the QFI, including temporal and mixed parameter-time components, has recently been proposed in Ref.~\cite{Diaz:2025fjb}.
While Ref.~\cite{Tang:2026xap} studied the post-quench QFI on a momentum-time $(k,t)$ manifold and Ref.~\cite{Diaz:2025fjb} formulated a general time-dependent QFI on a parameter-time manifold, the physical role of mixed parameter-time metric components in non-equilibrium many-body systems remains largely unexplored.
In particular, it is still unclear how the tangent direction generated by the real-time evolution is geometrically correlated with the tangent direction associated with a deformation of the post-quench Hamiltonian parameter.

Motivated by these developments, we investigate the information geometry of non-equilibrium quantum many-body dynamics using the QFI on an extended information manifold, whose coordinates are a controllable post-quench Hamiltonian parameter $h$ and real time $t$.
The purpose of the present study is to establish a simple and useful framework for characterizing non-equilibrium quantum states from an information-geometric perspective.
To this end, we focus on the mixed metric component $g_{ht}$ and on metric-volume-based quantities, which characterize the geometric coupling between the parameter deformation and the temporal evolution, including both the overall metric scale and their directional alignment.
As a minimal example, we employ the transverse-field Ising chain model, which possesses a well-understood quantum critical point and provides a clean setting for examining the proposed geometric quantities. 

A central novelty of the present study lies in the systematic separation of parameter sensitivity, parameter-time geometric coupling, directional alignment, and local metric volume for post-quench many-body states. In addition, we compare these geometric quantities with two widely used measures of non-equilibrium dynamics, the entanglement entropy and the Loschmidt echo, and quantify their correlations through Pearson correlation analysis.

This paper is organized as follows.
In the next section, we explain the formulation of the one-dimensional transverse-field Ising model as a benchmark model, and the geometric quantities such as the QFI metric and the metric-volume density (MVD).
Section\,\ref{sec:numerical_results} presents the numerical results, and Sec.\,\ref{sec:discussion} discusses the results.
Section\,\ref{sec:summary} is devoted to the summary.

\section{Formulation}
\label{sec:formulation}
In this section, we introduce the minimal model and the geometric quantities.
For simplicity, we employ the transverse-field Ising chain model.
The same construction can in principle be applied to other models.

\subsection{Transverse-field Ising chain model}
We employ the one-dimensional transverse-field Ising chain model~\cite{pfeuty1970one,pfeuty1971ising} with the open boundary condition;
\begin{align}
  H(h)
  &= - J \sum_{i=1}^{L-1} \sigma_i^z \sigma_{i+1}^z
     - h \sum_{i=1}^{L}\sigma_i^x,
  \label{eq:tfim_hamiltonian}
\end{align}
as the minimal model, where $L$ denotes the number of spins, $J > 0$ is the ferromagnetic nearest-neighbor coupling constant, $h$ is the transverse external magnetic field, and $\sigma_i^{x,z}$ denote Pauli operators on site $i$.
The model is one of the paradigmatic models of quantum critical phenomena and exhibits well-understood critical scaling behavior near the quantum critical point
~\cite{pfeuty1970one,pfeuty1971ising}.

In the thermodynamic limit, the model exhibits a quantum critical point at $h_c = J$~\cite{pfeuty1970one,pfeuty1971ising}.
For $h < J$, the system is in the ferromagnetic phase, while for $h > J$ it is in the paramagnetic phase. 
In this study, we set $J = 1$.
Throughout this study, energies and the transverse field $h$ are measured in units of $J$, and the evolution time $t$ in units of $1/J$; both $h$ and $t$ are therefore treated as dimensionless variables.

\subsection{Quantum quench protocol}
We prepare the ground state of an initial Hamiltonian as
\begin{align}
  \ket{\psi_0}
  &= \ket{ \psi_\mathrm{GS} (h_i)},
\end{align}
where $h_i$ denotes the initial magnetic field.
At $t = 0$, the magnetic field suddenly changes from $h_i$ to $h_f$, and the quantum state evolves according to
\begin{align}
  \ket{\psi(t;h_f,h_i)} &= e^{-iH(h_f)t} \ket{\psi_\mathrm{GS}(h_i)},
  \label{eq:quench_state}
\end{align}
where $\exp(-i H t)$ is the time-evolution operator.

\begin{figure}[t]
\centering
\begin{tikzpicture}[scale=1.0, >=stealth]
    \draw[->, thick] (-0.1,0) -- (5.2,0) node[right] {$h$};
    \draw[->, thick] (0.1,-0.2) -- (0.1,3.8) node[above] {$t$};
    \foreach \x in {0.5,1.2,1.9,2.6,3.3,4.0,4.7}{
    \draw[gray!45, thin] plot[smooth] coordinates { (\x,0.3) (\x+0.15,1.0) (\x-0.05,1.8) (\x+0.18,2.6) (\x,3.3) }; }
    \foreach \y in {0.4,1.0,1.6,2.2,2.8,3.4}{ \draw[gray!45, thin] plot[smooth] coordinates { (0.4,\y) (1.3,\y+0.15) (2.2,\y-0.05) (3.2,\y+0.12) (4.8,\y) }; }
    \coordinate (P) at (2.6,1.6);
    \coordinate (vt) at (1.35,0.35);
    \coordinate (vh) at (0.35,1.15);
    \filldraw[black] (P) circle (2pt);
    \node[above left] at (P) {$|\psi(h,t)\rangle$};
    \draw[fill=green!15, draw=green!15, opacity=0.7] (P) -- ($(P)+(vt)$) -- ($(P)+(vt)+(vh)$) -- ($(P)+(vh)$) -- cycle;;
    \draw[->, very thick, blue] (P) -- ++(1.35,0.35) node[right] {$\partial_h|\psi\rangle$};
    \draw[->, very thick, red] (P) -- ++(0.35,1.15) node[above] {$\partial_t|\psi\rangle$};
    \draw[thick] ($(P)+(0.45,0.12)$) arc[start angle=15,end angle=70,radius=0.5];
    \node at ($(P)+(0.5,0.5)$) {$\theta$};
\end{tikzpicture}
\caption{
Schematic picture of the information manifold $(h,t)$.
Each point represents the time-evolved state $|\psi(h,t)\rangle$.
The tangent vectors $\partial_h|\psi\rangle$ and $\partial_t|\psi\rangle$ define the local geometry of the manifold.
}
\label{fig:manifold}
\end{figure}
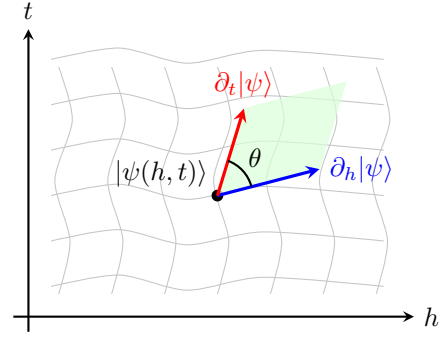

For the purpose of constructing the information metric, it is convenient to regard $h$ as the post-quench Hamiltonian parameter and define
\begin{align}
  \ket{\psi(h,t)} &= e^{-iH(h)t} \ket{\psi_\mathrm{GS}(h_i)}.
  \label{eq:psi_evolution}
\end{align}
The initial value $h_i$ is fixed, while $h$ and $t$ are treated as coordinates of the non-equilibrium information manifold; see Fig.~\ref{fig:manifold} for a schematic illustration of the manifold.

\subsection{Quantum geometric tensor and QFI metric}

For the coordinates $\lambda^\mu=(h,t)$, the neighboring states around a point $|\psi(h,t)\rangle$
are $|\psi(h+\delta h,t)\rangle$ and $|\psi(h,t+\delta t)\rangle$.
The corresponding tangent vectors $\partial_h|\psi\rangle$ and $\partial_t|\psi\rangle$ define the local geometry of the information manifold.
The geometric structure of the manifold of quantum states can be characterized by the quantum geometric tensor~\cite{Provost:1980nc}.
For a normalized pure state $\ket{\psi(\lambda)}$, the quantum geometric tensor is
\begin{align}
    G_{\mu\nu}
    &= \left\langle \partial_\mu \psi \middle|
                    \partial_\nu \psi \right\rangle
     - \left\langle \partial_\mu \psi \middle|
                    \psi \right\rangle
       \left\langle \psi \middle| \partial_\nu \psi \right\rangle. 
       \label{eq:G_mu_nu}
\end{align}
The real part of Eq.\,(\ref{eq:G_mu_nu}) defines the QFI metric $g_{\mu \nu}$~\cite{Braunstein:1994zz}
\begin{align}
  g_{\mu \nu} &= 4 \, \mathrm{Re} \, G_{\mu \nu},
\end{align}
while the imaginary part determines the Berry curvature~\cite{Berry:1984jv,Kolodrubetz:2017ofs}.

In the case with the single parameter $h$, the component $g_{hh}$ is the QFI associated with
changes in $h$; it is proportional to the fidelity susceptibility $\chi_\mathrm{F}$~\cite{Liu:2014mip,Liu:2019xfr}.
For two nearby states, $\ket{\psi(h)}$ and $\ket{\psi(h+\delta h)}$, the fidelity is expanded as
\begin{align}
  | \bra{\psi(h)} \ket{\psi(h+\delta h)} |^2
  &= 1 - \frac{1}{4} g_{hh} \, \delta h^2 + {\cal O}(\delta h^3),
\end{align}
where the term proportional to $\delta h$ vanishes due to the normalization $\langle \psi | \psi \rangle = 1$.
In this convention, $g_{hh} = 4 \chi_\mathrm{F}$.
Thus, a large value of $g_{hh}$ means that the state is highly sensitive to a small change of $h$.
Recent developments have emphasized the dynamical aspects of the QFI and its relation to quantum dynamics; see Ref.~\cite{Scandi:2023cjv} as a review.

\subsection{Information manifold in ($h,t$) space}
We now extend the information manifold by treating real-time $t$ as an additional coordinate and the corresponding metric
\begin{align}
  g_{\mu\nu}
  &= \begin{pmatrix}
        g_{hh} & g_{ht}\\
        g_{th} & g_{tt}
     \end{pmatrix}.
  \label{eq:g_mu_nu}
\end{align}
Since the QFI metric is symmetric, $g_{ht}=g_{th}$.
Note that we can consider higher-dimensional information manifolds if additional parameters are included, such as $(J,h,t)$.

Each component of Eq.\,(\ref{eq:g_mu_nu}) has a clear interpretation.
The component $g_{hh}$ measures the sensitivity of the time-evolved state to $h$.
The component $g_{tt}$ measures the distinguishability of neighboring states in $t$.
For unitary dynamics generated by a $t$-independent Hamiltonian, this component is related to the energy variance,
\begin{align}
    g_{tt}
    &= 4 \left( \langle H^2 \rangle - \langle H \rangle^2 \right)
     = 4 \, (\Delta H)^2.
     \label{eq:deltaH}
\end{align}
The mixed component $g_{ht}$ is the metric inner product between the parameter $h$ and temporal tangent directions. Thus, $g_{ht}=0$ corresponds to local orthogonality of the two directions. We refer to the non-orthogonality associated with nonzero $g_{ht}$ as a geometric coupling between the two tangent directions on the information manifold.

\begin{figure*}[t]
\centering
\includegraphics[width=0.32\linewidth]{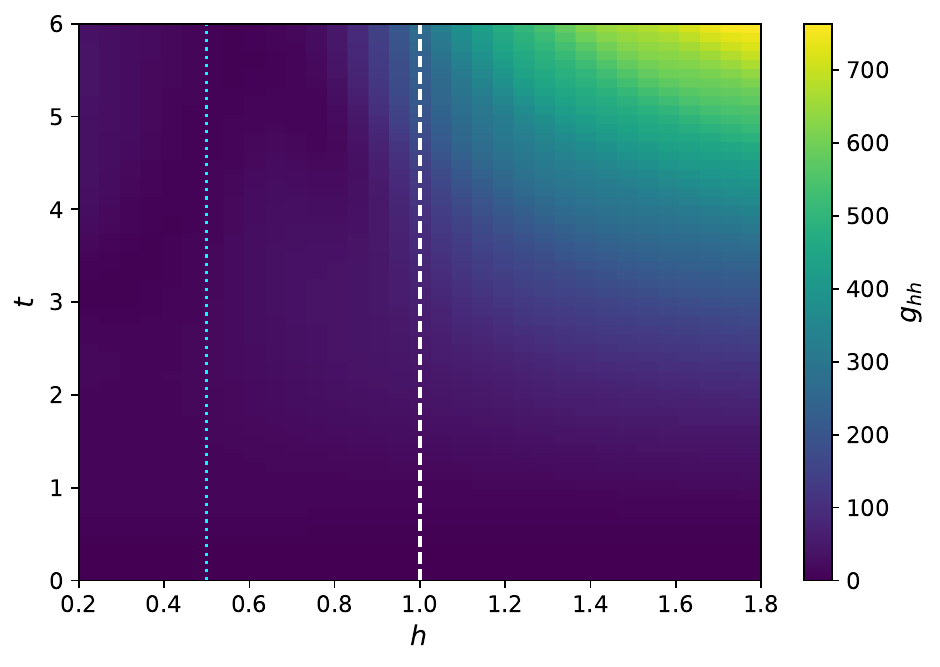}
\includegraphics[width=0.32\linewidth]{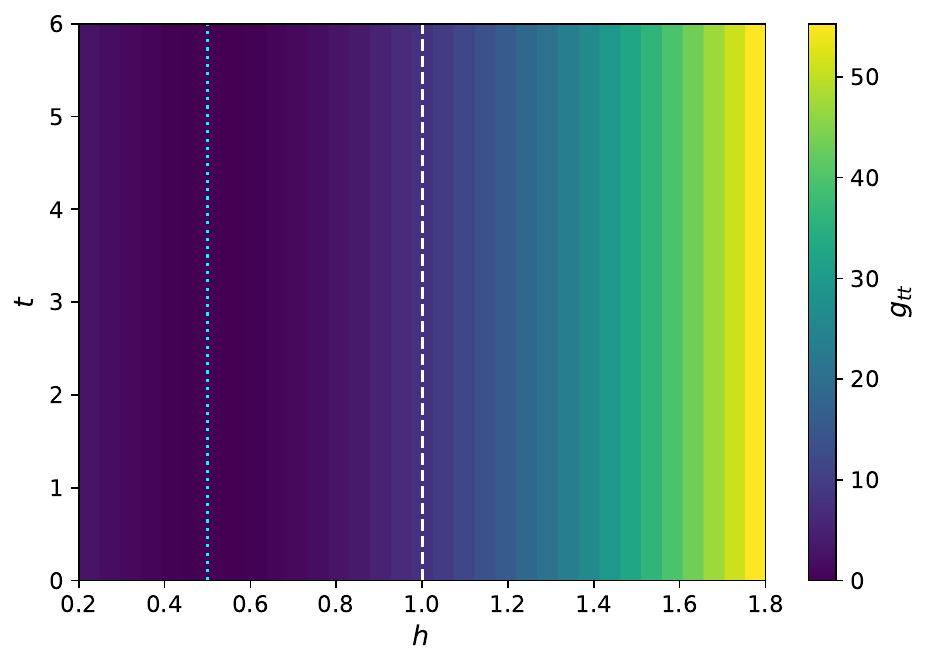}
\includegraphics[width=0.32\linewidth]{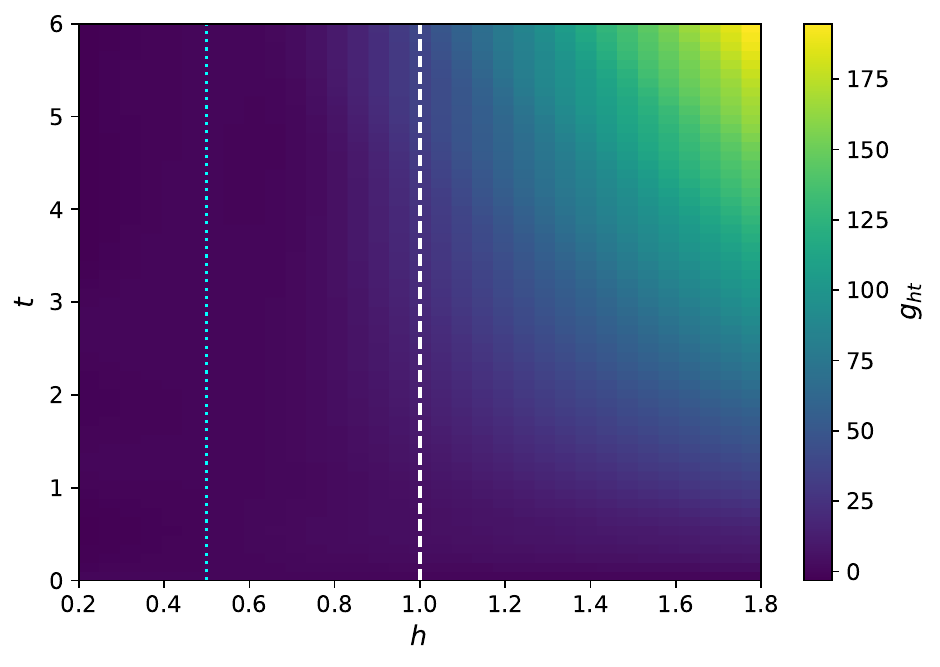}
\caption{
Heatmaps of the QFI metric components $g_{hh}$, $g_{tt}$, and $g_{ht}$ on the information manifold $(h,t)$.
The vertical dashed line indicates the equilibrium critical point $h=1$ in the thermodynamic limit, while the dotted line denotes the initial field $h=h_i=0.5$.
}
\label{fig:gmetric}
\end{figure*}

\subsection{Metric-volume density}
\label{sebsec:MVD}
Based on the QFI metric, we define the MVD as
\begin{align}
  D_\mathrm{MV}(\lambda) &= \sqrt{\det g_{\mu \nu} (\lambda)}.
  \label{eq:mvd_general}
\end{align}
In the two-dimensional case, it becomes
\begin{align}
  D_\mathrm{MV}(\lambda)
  &= \sqrt{ g_{hh}(\lambda) g_{tt}(\lambda) - g_{ht}^2(\lambda) },
  \label{eq:dmv}
\end{align}
Since $D_\mathrm{MV}$ is a metric density, it depends on the chosen coordinates on the manifold.
In the present study, we fix the natural dimensionless coordinates $h/J$ and $Jt$, whereas the normalized quantity $\widetilde D_\mathrm{MV}$ and the alignment coefficient $\rho_{ht}$, which are defined below, are used to characterize their alignment properties, independent of the choice of coordinate scale.

It is also useful to introduce the normalized alignment coefficient $\rho_{ht}$ as
\begin{align}
    \rho_{ht} &= \frac{g_{ht}}{\sqrt{g_{hh}g_{tt}}}.
    \label{eq:rho_ht}
\end{align}
We may interpret $\rho_{ht}$ as a measure of the angle between the two tangent directions; this quantity quantifies the geometric alignment between the $h$ direction and the $t$ direction.
The sign of $\rho_{ht}$ distinguishes the local orientation of the tangent vectors.

In terms of $\rho_{ht}$, the MVD can be written as
\begin{align}
  D_\mathrm{MV}(\lambda)
  &= \sqrt{g_{hh}(\lambda)} \sqrt{g_{tt}(\lambda)} \sqrt{1 - \rho_{ht}^2(\lambda)}.
  \label{eq:mvd_rho_relation}
\end{align}
It measures the local geometric area generated by the two independent tangent directions on the information manifold. This expression shows that $D_\mathrm{MV}$ results from a competition between two effects: the magnitude of $g_{hh}$ and $g_{tt}$, and the degree of independence between the two directions, $1-\rho_{ht}^2$.

To extract the alignment properties of two directions from $D_\mathrm{MV}$, we define the normalized MVD as
\begin{align}
    \widetilde D_\mathrm{MV}(\lambda)
    &= \frac{D_\mathrm{MV}(\lambda)}{\sqrt{g_{hh}(\lambda) g_{tt}(\lambda)}}
     = \sqrt{1 - \rho_{ht}^2(\lambda)}.
    \label{eq:mvd_normalized_rho}
\end{align}
It measures the degree of geometric independence between the parameter and the temporal direction.
The normalized MVD has a simple geometric meaning that it represents the normalized  local area element on the manifold $(h,t)$.

\subsection{Entanglement entropy and Loschmidt echo}

To clarify the physical meaning of the information-geometric quantities, we compare them with widely used measures of non-equilibrium dynamics: the entanglement entropy and the Loschmidt echo.
We consider the entanglement entropy, and the Loschmidt echo. 


The entanglement entropy for a subsystem $A$ is defined as
\begin{align}
  S_A(t)
  &= - \mathrm{Tr} \, \left[ \rho_A(t) \log \rho_A(t) \right],
\end{align}
where $\rho_A(t)$ is the reduced density matrix of the subsystem $A$.
The entanglement entropy $S_A$ quantifies the entanglement between subsystem $A$ and its complement, and its growth during the real-time evolution reflects the generation and spreading of entanglement. For details of the entanglement entropy, see Refs.~\cite{Calabrese:2005in,Vidal:2002rm}. 
In the numerical calculation, we set $A$ as one half of the chain and denote the corresponding entropy as $S_\mathrm{half}(t)$.

The Loschmidt amplitude ${\cal G}$ and the Loschmidt echo ${\cal L}$ are defined as
\begin{align}
  {\cal G}(t)
  &= \bra{\psi_0} e^{-iH(h_f)t} \ket{\psi_0},
\end{align}
and
\begin{align}
  {\cal L}(t) &= |{\cal G}(t)|^2.
\end{align}
The Loschmidt echo measures the overlap between the initial state and the time-evolved state, quantifying how closely the evolving state remains to the initial one.
These quantities are useful for diagnosing dynamical quantum phase transitions; see Refs.~\cite{Heyl:2013ywe,Heyl:2014qan}.

\begin{figure}[t]
\centering
\includegraphics[width=1.0\linewidth]{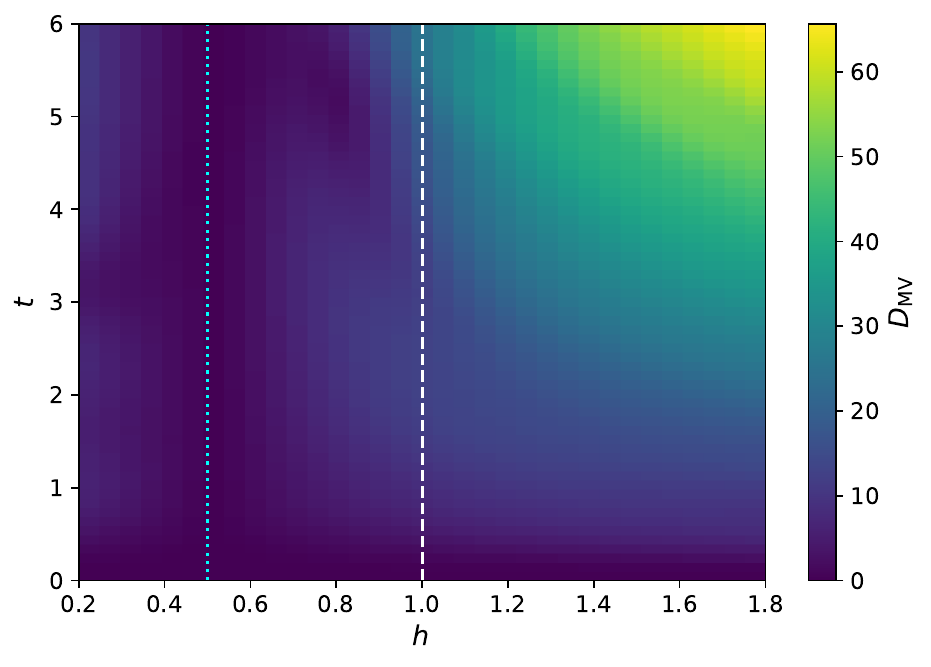}
\caption{
The MVD on the information manifold $(h,t)$.
The dashed and dotted lines are the same as in Fig.~\ref{fig:gmetric}. 
}
\label{fig:mvd_heatmap}
\end{figure}

\section{Numerical results}
\label{sec:numerical_results}
In this study, we evaluate the information geometry of the transverse-field Ising chain model using exact diagonalization.
The number of spins is fixed to $L=8$.
The initial state is chosen as the ground state with $h_i = 0.5$.
For each post-quench field $h$, the time-evolved state is numerically constructed using Eq.\,(\ref{eq:psi_evolution}).
The post-quench field $h$ and real time $t$ are scanned on a rectangular grid, $h \in [0.2,1.8],~t \in [0,6]$, with $33$ field points and $61$ time points.
The derivative in $t$ is analytically evaluated using $\partial_t|\psi\rangle=-iH|\psi\rangle$, while the derivative in $h$ is evaluated by a central finite difference.
For the initial state with $h_i=0.5$ considered here, the enhancement occurs on the large-$h$ side.
Because the post-quench field, the quench amplitude $|h-h_i|$, and the equilibrium phase are varied simultaneously, the present result should not be interpreted as a phase-only effect.
The dependence on the initial state will be studied elsewhere.

\begin{figure}[t]
\centering
\includegraphics[width=1.0\linewidth]{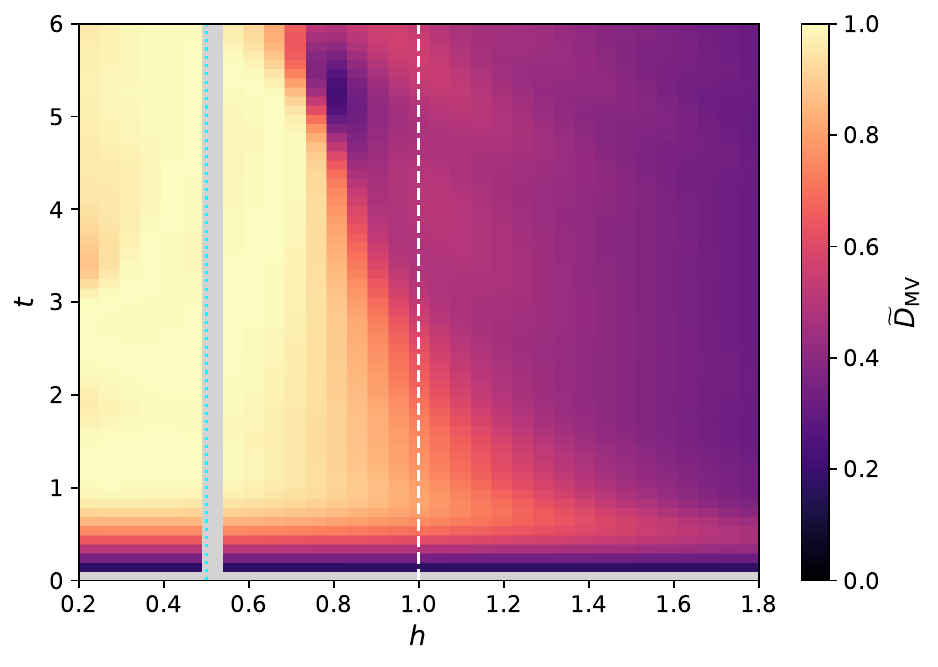}
\caption{
The normalized MVD on the information manifold $(h,t)$.
The gray region indicates masked areas where the normalization becomes ill-defined because of the no-quench condition. The dashed and dotted lines are the same as in Fig.~\ref{fig:gmetric}. 
}
\label{fig:mvd_norm_heatmap}
\end{figure}
\begin{figure}[t]
\centering
\includegraphics[width=1.0\linewidth]{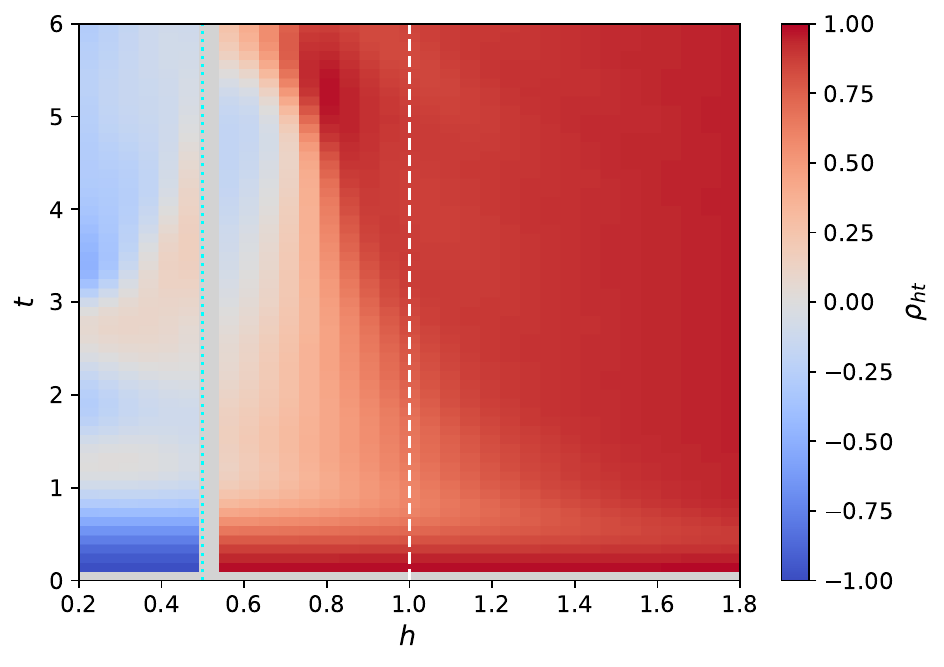}
\caption{The normalized alignment
coefficient $\rho_{ht}$ on the information manifold $(h,t)$.
The gray region indicates masked areas where the normalization becomes ill-defined because of the no-quench condition.
The dashed and dotted lines are the same as in Fig.~\ref{fig:gmetric}.
}
\label{fig:rhoht_heatmap}
\end{figure}

Figure~\ref{fig:gmetric} shows the QFI metric components $g_{hh}$, $g_{tt}$, and $g_{ht}$ on the information manifold $(h,t)$.
The component $g_{hh}$ exhibits a strong enhancement in the paramagnetic region ($h>1$), indicating a high sensitivity of the evolved state to changes in $h$.
By contrast, $g_{tt}$ appears constant in time, as expected from Eq.~\eqref{eq:deltaH}, which implies its invariance under unitary evolution for a $t$-independent Hamiltonian.
The mixed component $g_{ht}$ becomes large in the paramagnetic region.
This behavior indicates that the parameter and temporal directions are not independent on the information manifold.

Figure~\ref{fig:mvd_heatmap} shows the MVD on the information manifold $(h,t)$.
A pronounced enhancement is observed in the paramagnetic region ($h>1$), where both $g_{hh}$ and $g_{ht}$ become large.
The MVD combines the magnitudes of $g_{hh}$ and $g_{tt}$ with their geometric coupling $g_{ht}$. It thus encodes both the overall metric scale and the local two-dimensional distinguishability of the information manifold.

To remove the overall scale associated with the diagonal metric components ($\sqrt{g_{hh}g_{tt}}$), we consider the normalized quantity defined in Eq.~(\ref{eq:mvd_normalized_rho}).
According to Eq.~(\ref{eq:mvd_normalized_rho}), this quantity measures the geometric independence between the $h$ and $t$ directions.
The result is shown in Fig.~\ref{fig:mvd_norm_heatmap}. Unlike the MVD, the normalized MVD is enhanced predominantly on the small-$h$ side rather than in the paramagnetic region. This indicates that the strong large-$h$ enhancement of the MVD is mainly induced by the overall scale $\sqrt{g_{hh}g_{tt}}$. Note that a gray region appears in Fig.~\ref{fig:mvd_norm_heatmap} at $h=h_i=0.5$, corresponding to the no-quench line: at the no-quench line, the initial state is an eigenstate of the Hamiltonian with $h=h_i$, and the time evolution changes only its overall phase, so that $g_{tt}=0$.
Consequently, the metric becomes rank deficient and the normalized alignment quantities are undefined on this line.

\begin{figure}[t]
\centering
\includegraphics[width=1.0\linewidth]{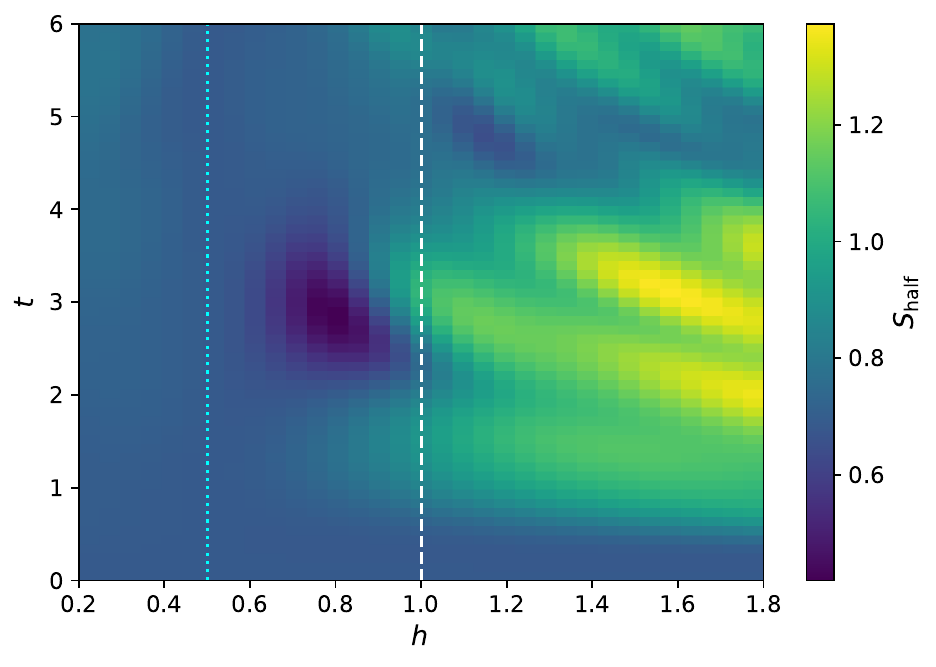}
\caption{
The half-chain entanglement entropy on the information manifold $(h,t)$.
The dashed and dotted lines are the same as in Fig.~\ref{fig:gmetric}. 
}
\label{fig:entropy_heatmap}
\end{figure}

The $\rho_{ht}$ on the information manifold $(h,t)$ is also shown in Fig.~\ref{fig:rhoht_heatmap}.
According to Eq.~(\ref{eq:mvd_normalized_rho}), $|\rho_{ht}|$ measures the degree of geometric alignment between the $h$ and $t$ directions, complementary to the geometric independence measured by $\widetilde D_\mathrm{MV}$.
Large values of $|\rho_{ht}|$ appear primarily in the high $h$ region, indicating a strong alignment between the parameter and the temporal direction, consistent with the behavior of $\widetilde D_\mathrm{MV}$ shown in Fig.~\ref{fig:mvd_norm_heatmap}.

Figure~\ref{fig:entropy_heatmap} shows the half-chain entanglement entropy $S_\mathrm{half}$. The entropy increases rapidly after the quench and exhibits oscillatory structures during the real-time evolution. Although both the half-chain entanglement entropy and the MVD are found to be enhanced in the paramagnetic region, their detailed patterns are different. It is also found that the region of maximal half-chain entanglement entropy does not coincide with the region where $\widetilde D_\mathrm{MV}$ becomes small, equivalently where $|\rho_{ht}|$ becomes large.

\begin{figure}[t]
\centering
\includegraphics[width=1.0\linewidth]{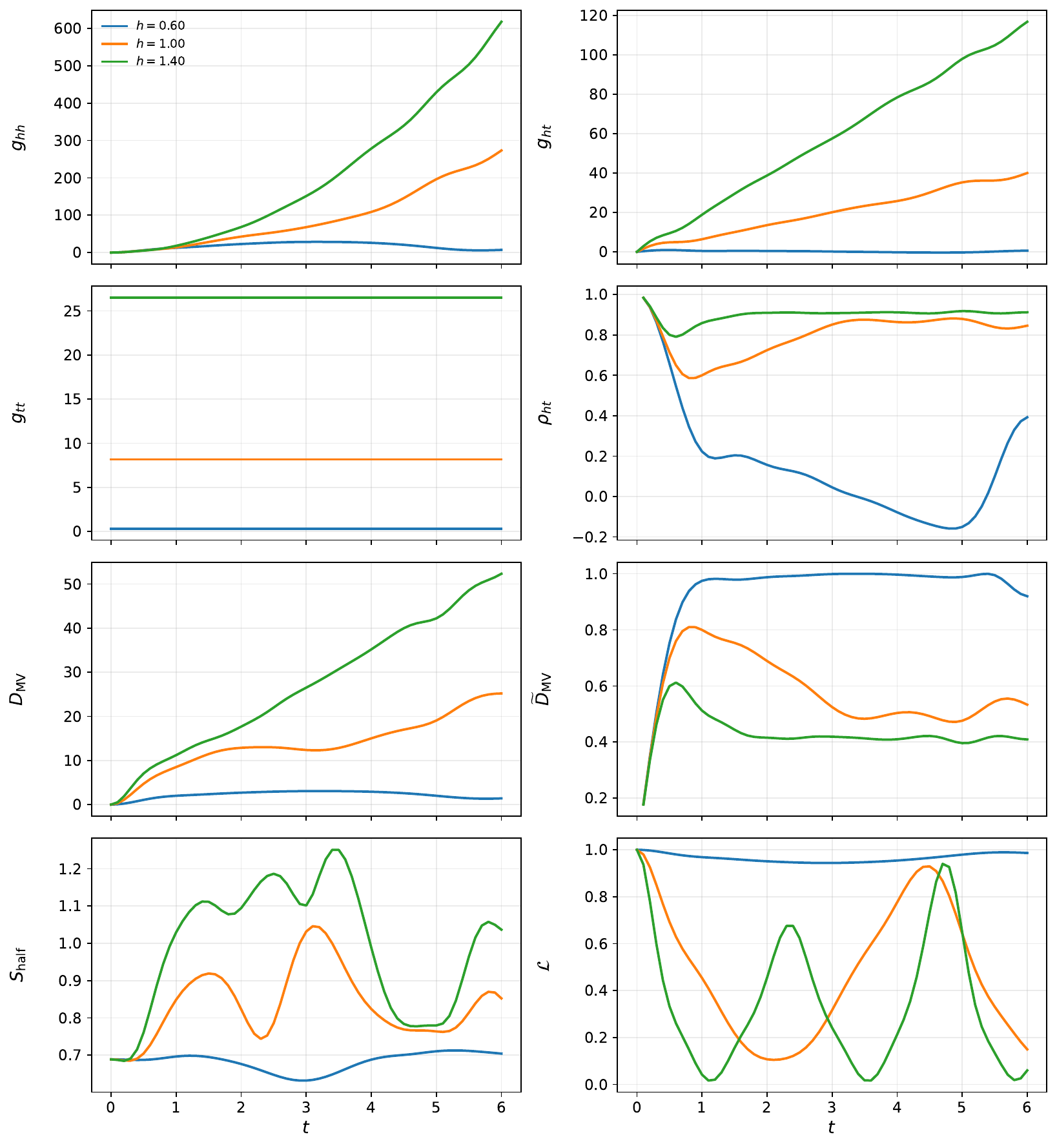}
\caption{
The real-time evolution of $D_\mathrm{MV}$, $\widetilde{D}_\mathrm{MV}$, $\rho_{ht}$, $g_{hh}$, $S_\mathrm{half}$, and ${\cal L}$ for $h=0.6$, $1.0$, and $1.4$.
These values correspond to quenches in the ferromagnetic phase, near the equilibrium critical point, and in the paramagnetic phase, respectively.
}
\label{fig:timecuts}
\end{figure}
\begin{figure}[t]
\centering
\includegraphics[width=0.9\linewidth]{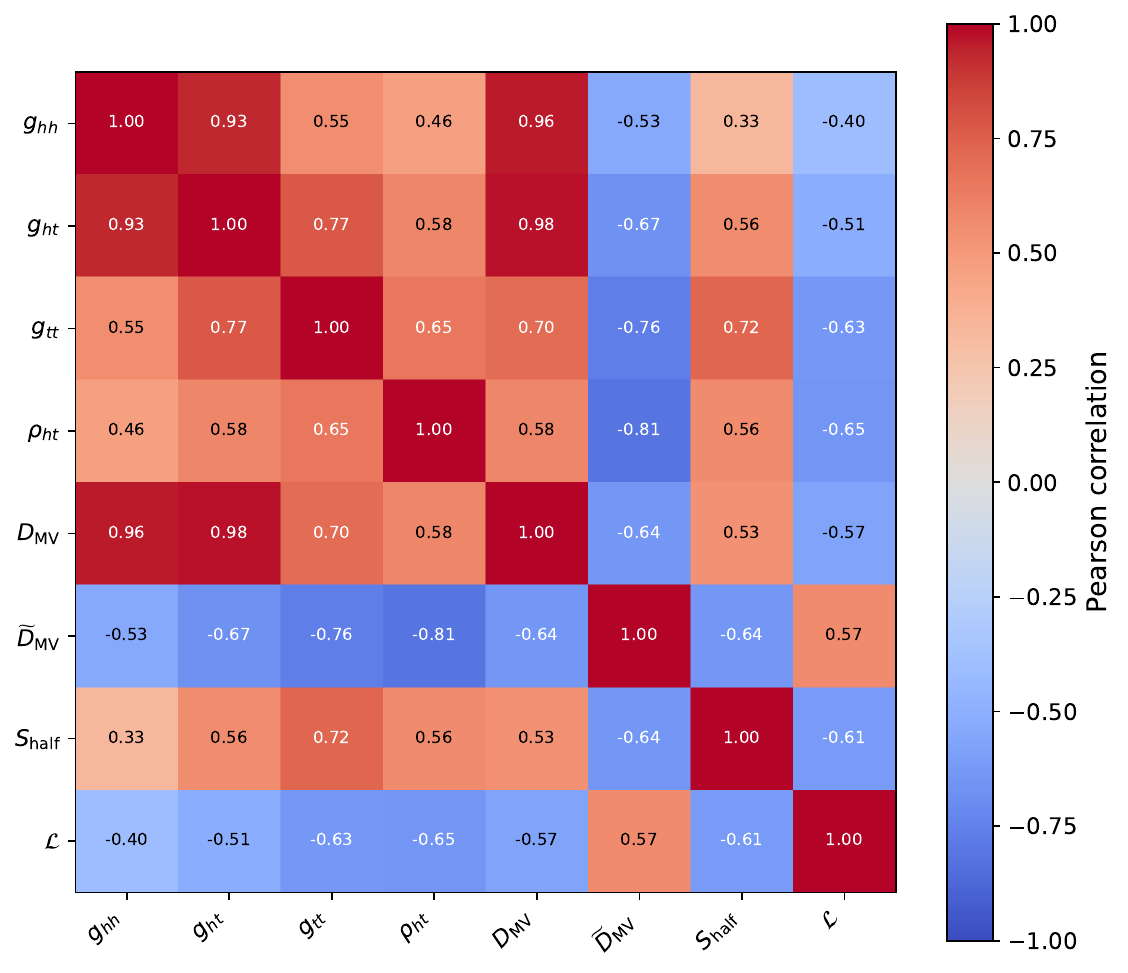}
\caption{
Descriptive Pearson correlation matrix among geometric and conventional observables over the $(h,t)$ grid.
Part of the correlation among the geometric quantities follows from their algebraic definitions.
The matrix is used as an overall descriptive summary and should not be interpreted as an inference based on independent samples. Direction-decomposed correlations are shown in Appendix~\ref{sec:correlations}.
}
\label{fig:corr_matrix}
\end{figure}

To investigate the real-time evolution in more detail, Fig.~\ref{fig:timecuts} shows the $t$ dependence of $D_\mathrm{MV}$, $\widetilde{D}_\mathrm{MV}$, $\rho_{ht}$, $g_{hh}$, $S_\mathrm{half}$, and ${\cal L}$ for $h=0.6$, $1.0$, and $1.4$, corresponding to quenches in the ferromagnetic phase, near the equilibrium critical point, and in the paramagnetic phase, respectively.
The MVD increases rapidly at early times and then saturates.
By contrast, the half-chain entanglement entropy and the Loschmidt echo exhibit pronounced oscillations.
This indicates that the MVD on the information manifold captures an aspect of the non-equilibrium dynamics distinct from both the generation and spreading of entanglement reflected in the half-chain entanglement entropy and the proximity of the evolving state to the initial state measured by the Loschmidt echo.

To summarize the relationships among the various observables, we compute Pearson correlation coefficients~\cite{pearson1895note} over the full $(h,t)$ dataset, after excluding grid points where the normalized geometric quantities are mathematically undefined because of the rank deficiency.
The correlation matrix is shown in Fig.~\ref{fig:corr_matrix}.
Note that we evaluate the correlations without no-quench value of $h$ above the reference time to avoid the numerical singularities.
Although substantial correlations are found between the geometric quantities and conventional observables, none of the correlation coefficients reaches $1$.
This indicates that no geometric quantity is simply related to either of the conventional observables through a perfect linear correspondence. Note that the correlation matrix is used only as a descriptive overview:
because the data form smooth deterministic trajectories on the $(h,t)$ grid, the matrix
should not be interpreted as an inference based on independent samples.
The directional coefficients in Appendix~\ref{sec:correlations} are used to distinguish temporal co-variation from post-quench-field dependence.

The most distinctive feature of the present framework is the emergence of the mixed geometric component $g_{ht}$, which quantifies the interplay between parameter variation and real-time evolution.
The MVD incorporates information from the mixed component and therefore probes aspects of the non-equilibrium dynamics beyond conventional observables.
Note that, in addition to $L=8$, we evaluate the geometric quantities for $L=12$ and $16$ and verify that the main geometric features are robust against finite-size effects, as shown in Appendix~\ref{sec:system_volume}.

\section{Discussion}
\label{sec:discussion}
In this section, we discuss the information manifold in a deeper sense and provide an extension to the thermal mixed state.

\subsection{Meaning of geometric quantities}
A distinctive aspect of the present study is the systematic investigation of the mixed metric component $g_{ht}$ from several complementary perspectives within a concrete many-body model, the transverse-field Ising chain.
Unlike the conventional fidelity-susceptibility component $g_{hh}$, $g_{ht}$ characterizes the interplay between parameter deformation and real-time evolution.

From a geometric point of view, the vectors $\partial_h \ket{\psi(h,t)}$ and $\partial_t \ket{\psi(h,t)}$ represent two tangent directions on the information manifold $(h,t)$.
The mixed metric component $g_{ht}$ quantifies the overlap between these directions.
The numerical results show that $g_{ht}$ is significantly enhanced in the large-$h$ region.
This observation suggests that, in this regime, the response of the quantum state to changes in $h$ cannot be completely separated from the intrinsic non-equilibrium dynamics.
A large $|g_{ht}|$ indicates a strong mixed metric coupling between the state variations induced by changing $h$ and by real-time evolution.

While $|\rho_{ht}|$ measures the degree of geometric alignment between the two tangent directions, the normalized MVD measures their geometric independence. These quantities contain the same information and provide complementary characterizations of the same local geometric relation. From an operational point of view, a large value of $g_{ht}$ indicates a strong overlap between state variations generated by parameter deformation and those generated by real-time evolution.

While $g_{tt}$ is directly related to the energy variance (\ref{eq:deltaH}) and to the dynamical speed on the manifold~\cite{Anandan:1990fq}, the mixed component $g_{ht}$ may be interpreted as a geometric coupling between the dynamical evolution of the state and its sensitivity to parameter deformation.
For a quantum state trajectory $| \psi(h,t) \rangle$, the QFI metric defines the infinitesimal line element as
\begin{align}
    ds^2 &= g_{hh} \, dh^2 + 2 g_{ht} \, dh \, dt + g_{tt} \, dt^2.
\end{align}
Along an arbitrary path $h=h(t)$, the geometric speed $v_\mathrm{G}$ is obtained as
\begin{align}
    v_\mathrm{G} &= \frac{ds}{dt} = \sqrt{ g_{tt} + 2g_{ht}\dot{h} + g_{hh} \dot{h}^2},
\end{align}
where $\dot h$ means $dh/dt$.
For the present sudden-quench dynamics, the post-quench parameter $h$ is fixed and therefore $\dot h = 0$.
In this case, the geometric speed reduces to
\begin{align}
    v_\mathrm{G} &= \sqrt{g_{tt}} = 2 \, \Delta H,
\end{align}
which is directly determined by the energy uncertainty entering the Mandelstam--Tamm quantum speed limit~\cite{1370285709897301766,Anandan:1990fq,Deffner:2017cxz}.
Thus, the component $g_{tt}$ quantifies how rapidly the quantum state moves on the manifold.
In contrast, the mixed metric component $g_{ht}$ does not directly determine the speed of real-time evolution.

The geometric meaning of $g_{ht}$ can be clarified by expressing the tangent vectors in terms of physical operators.
For the non-equilibrium state (\ref{eq:psi_evolution}), the $t$ derivative is given by
\begin{align}
    \partial_t |\psi(h,t)\rangle &= -iH(h)\, |\psi(h,t)\rangle .
\end{align}
The derivative of the time-evolution operator with respect
to $h$ is given by the standard parameter-derivative for operator exponentials~\cite{Wilcox:1967zz} as
\begin{align}
    \partial_h U(h,t) &= -i U(h,t) \mathcal A_h(t),
\end{align}
where $U(h,t) = e^{-iH(h)t}$ and
\begin{align}
    \mathcal A_h(t)
    &= \int_0^t ds\, U^\dag(h,s) \Bigl( \partial_h H(h) \Bigr) U(h,s).
    \label{eq:Ah}
\end{align}
Because $\partial_hH$ is Hermitian, $\mathcal A_h(t)$ is also Hermitian. The $h$ derivative of the state is therefore
\begin{align}
    |\partial_h\psi(h,t)\rangle
    &= -i e^{-iH(h)t} \mathcal A_h(t)|\psi_\mathrm{GS} \rangle .
    \label{eq:dh}
\end{align}
Similarly, the $t$ derivative is
\begin{align}
    | \partial_t\psi(h,t) \rangle
    &= -iH(h) | \psi(h,t) \rangle
    \nonumber\\
    &= -i U(h,t) H(h)| \psi_\mathrm{GS} \rangle ,
    \label{eq:dt}
\end{align}
where we used $[H,U]=0$.
Since $U(h,t)$ is unitary, it preserves inner products, $U^\dagger U=1$. Therefore,
Eqs.~\eqref{eq:dh} and \eqref{eq:dt} give
\begin{align}
    \langle \partial_h \psi | \partial_h \psi \rangle
    &= \langle \mathcal{A}_h^2(t) \rangle,
    \\
    \langle \partial_t \psi | \partial_t \psi \rangle
    &= \langle H^2 \rangle,
    \\
    \langle \partial_h \psi | \partial_t \psi \rangle
    &= \langle \mathcal{A}_h(t) H \rangle.
    \label{eq:tangent_inner_products}
\end{align}
and
\begin{align}
    \langle \partial_h \psi | \psi \rangle
    &= i \langle \mathcal A_h(t) \rangle,
    \\
    \langle \psi | \partial_h \psi \rangle
    &= -i \langle \mathcal{A}_h(t) \rangle,
    \\
    \langle \partial_t \psi | \psi \rangle
    &= i \langle H \rangle,
    \\
    \langle \psi | \partial_t \psi \rangle
    &= -i \langle H \rangle.
    \label{eq:parallel_components}
\end{align}
Substituting these expressions into the quantum geometric tensor (\ref{eq:G_mu_nu}), we have
\begin{align}
 g_{hh}&= 4 \Bigl[ \langle \mathcal{A}_h^2(t) \rangle
         - \langle \mathcal{A}_h(t) \rangle^2 \Bigr]
        = 4 \mathrm{Var} [\mathcal{A}_h(t)],
 \label{eq:ghh2}\\
 g_{tt} &= 4 \Bigl[ \langle H^2 \rangle - \langle H \rangle^2 \Bigr]
         = 4 \mathrm{Var} [H].
\end{align}
and
\begin{align}
    g_{ht} &= 4 \, \mathrm{Re} \Bigl[ \langle \mathcal{A}_h(t) H \rangle
            - \langle \mathcal{A}_h(t) \rangle
              \langle H \rangle \Bigl].
    \label{eq:ght2}
\end{align}
Introducing the fluctuation operators
\begin{align}
    \delta \mathcal{A}_h(t)
    &= \mathcal{A}_h(t) - \langle \mathcal{A}_h(t) \rangle,
    \nonumber\\
    \delta H &= H - \langle H \rangle,
\end{align}
we obtain
\begin{align}
    g_{ht}
    &= 4 \, \mathrm{Re} \langle \delta\mathcal A_h(t) \, \delta H \rangle
     = 2 \, \Bigl\langle \left\{ \delta\mathcal A_h(t), \delta H \right\} \Bigr\rangle.
\end{align}
When the symmetrized covariance is defined by
\begin{align}
    \mathrm{Cov} (X,Y)
    &=\frac{1}{2} \Bigl\langle
      \left\{ X - \langle X \rangle,
              Y - \langle Y \rangle
      \right\} \Bigr\rangle,
\end{align}
$g_{ht}$ takes the compact form as
\begin{align}
 g_{ht} &= 4 \, \mathrm{Cov} ( \mathcal{A}_h(t), H ).
 \label{eq:ght}
\end{align}
Thus, $g_{ht}$ measures the cross covariance between the integrated generalized force $\mathcal A_h(t)$ and the Hamiltonian generating the real-time evolution.
It therefore quantifies whether the state deformation induced by a variation of the post-quench parameter $h$ is coupled to the temporal dynamical flow. In the short-time limit, $\mathcal A_h(t)\simeq t\partial_h H$, yielding $g_{ht} \simeq 4t\,\mathrm{Cov}(\partial_h H,H),$ which directly relates the mixed metric to the covariance between the generalized force and the energy. In the large-$h$ regime, the transverse-field term dominates the Hamiltonian.
Consequently, the deformation generator $\partial_h H$ and the Hamiltonian itself become increasingly aligned.
This naturally enhances the covariance between $\mathcal A_h$ and $H$, leading to the observed increase of $g_{ht}$.

An important aspect of $g_{ht}$ is its potential experimental accessibility. Unlike a purely geometric quantity defined only in Hilbert space, $g_{ht}$ expressed in Eq.~(\ref{eq:ght}) is expressed in terms of dynamical correlation functions.
The mixed metric can be interpreted as the covariance between the energy and the accumulated response associated with the control parameter $h$.
For the transverse-field Ising model, $\mathcal A_h(t)$ corresponds to the time-integrated transverse magnetization operator.
Consequently, $g_{ht}$ characterizes the correlation between the energy fluctuation and the integrated transverse-spin response generated during the non-equilibrium evolution.

The operator $\mathcal A_h(t)$, which enters $g_{ht}$, is the time integral of the generalized force $\partial_h H$ in the Heisenberg picture, as given in Eq.~(\ref{eq:Ah}). Introducing the susceptibility based on the Kubo formula~\cite{kubo1957statistical,Kolodrubetz:2017ofs} as
\begin{align}
    \chi_{hh}(t)
    &= i\theta(t) \Bigl\langle
       \left[ \partial_h H(t), \partial_h H(0) \right] \Bigr\rangle,
\end{align}
the generalized force $\partial_h H$ can be viewed as the perturbation operator entering the linear-response function.
Since the same generalized force also enters $\mathcal A_h(t)$, $g_{ht}$ is naturally expected to be related to the non-equilibrium response associated with variations of $h$.

\subsection{Application to non-integrable dynamics}
The present information-geometric framework can be naturally applied to non-integrable quantum dynamics. Integrability is closely related to the distinction between regular and chaotic dynamics in both classical and quantum systems~\cite{DAlessio:2015qtq}. As an example, we introduce an integrability-breaking parameter $\gamma$ and consider the extended information manifold with coordinates $(\gamma,t)$, regarded as a submanifold of the more general manifold $(h,\gamma,t,\cdots)$.

The QFI is defined as
\begin{align}
    g_{\mu\nu} =
    \begin{pmatrix}
        g_{\gamma\gamma} & g_{\gamma t}\\ g_{t\gamma} & g_{tt} 
    \end{pmatrix},
\end{align}
where $g_{\gamma\gamma}$ characterizes the sensitivity of the quantum state to integrability-breaking perturbations and $g_{\gamma t}$ describes the coupling between real-time evolution and departures from integrability.
In analogy with the $(h,t)$ manifold, one may define the alignment coefficient as
\begin{align}
    \rho_{\gamma t}
    &= \frac{g_{\gamma t}} {\sqrt{g_{\gamma\gamma}g_{tt}}}, 
\end{align}
and the MVD as
\begin{align}
    D_{\rm MV}
    &= \sqrt{ g_{\gamma\gamma}g_{tt} - g_{\gamma t}^2 }.
\end{align}
These quantities provide geometric measures of the response of the quantum state to integrability-breaking perturbations.
The geometric structure of the ($\gamma,t$) manifold may therefore provide a useful probe of the crossover from integrable to non-integrable dynamics. A detailed investigation of the resulting geometric structures is beyond the scope of the present study.

\begin{table*}[t]
\caption{Summary of the geometric observables and their interpretations.}
\begin{ruledtabular}
    \begin{tabular}{lll}
    Observable & Measures & Interpretation
    \\
    \hline
    $g_{tt}$ & How rapidly the state changes in $t$ & Geometric speed
    \\
    $g_{hh}$ & How sensitive the state is to $h$ & Geometric sensitivity
    \\
    $g_{ht}$ & How strongly the two directions are coupled & Geometric cross covariance
    \\
    $\rho_{ht}$ & How the two directions align & Geometric alignment
    \\
    $\widetilde D_\mathrm{MV}$ & How independent the two directions are & Geometric independence
    \\
    \end{tabular}
\end{ruledtabular}
\label{table:summary}
\end{table*}

\subsection{Extension to thermal mixed state with quench in closed system}
An important extension is the application to thermal mixed states.
For the closed thermal system, the pure state can be replaced by a density matrix
\begin{align}
    \varrho(h,t,T)
    &= e^{-iH(h)t} \, \varrho_0(T) \, e^{iH(h)t},
\end{align}
where
\begin{align}
    \varrho_0(T) &= \frac{e^{-\beta H(h_i)}}{{\cal Z}}.
\end{align}
Here ${\cal Z}$ denotes the partition function and $\beta$ is the inverse temperature $\beta = 1/T$.
Note that the present temperature characterizes the thermal preparation of the initial state rather than the temperature of an external environment.
The real-time evolution is assumed to be unitary, and therefore the system remains closed throughout the dynamics; we here consider the quench in a closed system.
In this sense, the present extension does not consider the dissipative thermalization in an open quantum system.
The natural coordinates of the extended information manifold then become
\begin{align}
    \lambda^\mu = (h,t,T),
\end{align}
leading to a three-dimensional mixed-state information manifold.
Note that we should use dimensionless coordinates as explained in Sec.\,\ref{sebsec:MVD}.

For mixed states, one natural construction of the information geometry is based on the Bures metric~\cite{bures1969extension} or, equivalently, the QFI metric~\cite{Braunstein:1994zz,Paris:2008zgg} up to a conventional factor $4$.
The metric tensor becomes
\begin{align}
    g_{\mu\nu} &=
    \begin{pmatrix}
        g_{hh} & g_{ht} & g_{hT}
        \\
        g_{th} & g_{tt} & g_{tT}
        \\
        g_{Th} & g_{Tt} & g_{TT}
    \end{pmatrix},
\end{align}
where the new mixed components $g_{hT}$ and $g_{tT}$ characterize the geometric coupling between the direction of initial temperature and the directions of parameter deformation and time evolution, respectively.
Here, we assume non-degenerate case.

The MVD can also be generalized straightforwardly.
For the three-dimensional manifold $(h,t,T)$, it is natural to define
\begin{align}
    D_\mathrm{MV}
    &= \sqrt{\det g}
    \nonumber\\
    &= \Bigl[ g_{hh} g_{tt} g_{TT} + 2 g_{ht} g_{hT} g_{tT}
    \nonumber\\
    & ~~~~~~~~~
            - g_{hh} g_{tT}^{2} - g_{tt} g_{hT}^{2} - g_{TT} g_{ht}^{2}
        \Bigr]^{\frac{1}{2}}.
\end{align}
This quantity measures the local geometric volume generated by parameter deformation, real-time evolution, and thermal fluctuations.
Its magnitude contains both the metric scales of the three directions and the extent to which these directions remain linearly independent.

The normalized MVD then become
\begin{align}
    \widetilde{D}_\mathrm{MV}
    &= \sqrt{ 1 -\rho_{ht}^{2} - \rho_{hT}^{2} - \rho_{tT}^{2} + 2\rho_{ht}\rho_{hT}\rho_{tT}},
\end{align}
where the alignment coefficients are defined as
\begin{align}
    \rho_{ht} &= \frac{g_{ht}}{\sqrt{g_{hh} g_{tt}}},
    \\
    \rho_{hT} &= \frac{g_{hT}}{\sqrt{g_{hh} g_{TT}}},
    \\
    \rho_{tT} &= \frac{g_{tT}}{\sqrt{g_{tt} g_{TT}}}.
\end{align}
The quantity $\rho_{tT}$ characterizes the scale-independent geometric alignment between the temporal direction and the direction associated with changes in the initial thermal preparation.
When $|\rho_{tT}|$ approaches $1$, the two directions become parallel on the extended information manifold.


An interesting future direction is to further explore the geometric properties of quantum-state trajectories on the information manifold and to investigate whether they are related to quantum chaos in non-integrable quantum many-body systems. Another further interesting extension is to consider the open quantum system coupled to a thermal environment.
In such a case, the density matrix may evolve according to, for example, the Lindblad equation~\cite{Gorini:1975nb,Lindblad:1975ef}.
The present framework can also be further extended by introducing an integrability-breaking coordinate.

\section{Summary}
\label{sec:summary}
\begin{figure}[t]
\centering
\begin{tikzpicture}[scale=1.2, >=stealth]
    \fill[green!15, opacity=0.7] (0,0) -- (2.8,0.8) -- (4.0,3.0) -- (1.2,2.2) -- cycle;
    \draw[dashed] (2.8,0.8) -- (4.0,3.0); 
    \draw[dashed] (1.2,2.2) -- (4.0,3.0);
    \draw[-> ,very thick,blue] (0,0) -- ++ (2.8,0.8); 
    \draw[-> ,very thick,red] (0,0) -- ++ (1.2,2.2);
    \draw (0.5,0.15) arc (0:53:0.5);
    \node[blue] at (1.8,0.1) {$\partial_t|\psi\rangle$};
    \node[left,red] at (0.45,1.2) {$\partial_h|\psi\rangle$};
    \node[green!50!black] at (2,1.5) {$\widetilde D_\mathrm{MV}$};
    \node[black] at (1.3,0.85) {$\rho_{ht} = \cos \theta$};    
    \node at (0.6,0.45) {$\theta$};
    \draw[blue,thick,fill=white] (2.9,0.85) circle (0.17);
    \node[blue] at (2.9,0.85) {\scriptsize $v_\mathrm{G}$};
\end{tikzpicture}
\caption{
Schematic picture of $v_\mathrm{G}$, $\rho_{ht}$ and $\widetilde{D}_\mathrm{MV}$ on the information manifold.
The length of the temporal tangent vector is quantified by $v_\mathrm{G} = \sqrt{g_{tt}}$ denoted by circle.
}
\label{fig:schematic_three}
\end{figure}
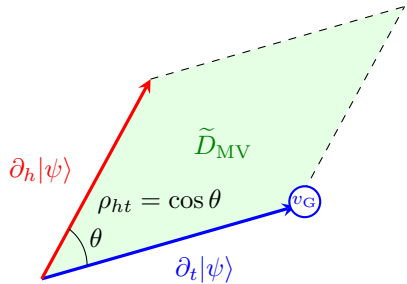

We have proposed an information-geometric description of non-equilibrium quantum dynamics based on the extended information manifold $(h,t)$.
The central quantity of the formulation is the mixed quantum Fisher information (QFI) metric component $g_{ht}$, which characterizes the interplay between parameter deformation and real-time evolution. The diagonal components $g_{tt}$ and $g_{hh}$, which characterize the speed of temporal evolution and parameter sensitivity, respectively, are also considered.
To distinguish the overall metric scale from the directional relation between parameter deformation and temporal evolution, we introduced the metric-volume density (MVD) $D_\mathrm{MV}$ and its normalized one $\widetilde{D}_\mathrm{MV}$ as quantities characterizing how the parameter and the temporal direction jointly span the manifold of distinguishable quantum states.
In addition, we introduced the alignment coefficient $\rho_{ht}$ which measures the angle between
the two tangent directions on the manifold. 
On the two dimensional manifold, $\rho_{ht}$ and $\widetilde D_{\rm MV}$ contain equivalent information about the magnitude of the alignment, but provide complementary geometric representations: $\rho_{ht}$ retains the orientation and the sign of the alignment, whereas $\widetilde D_{\rm MV}$ expresses the geometric independence.
We summarize our geometric insights in Table~\ref{table:summary} and our interpretation in Fig.~\ref{fig:schematic_three}.

For the initial state and quench protocol considered here, the diagonal component $g_{hh}$, which characterizes parameter sensitivity, and the mixed component $g_{ht}$ are strongly enhanced on the large-$h$ side of the parameter scan, which lies in the
paramagnetic regime of the thermodynamic model. The MVD $D_{\rm MV}$ is correspondingly enhanced in the large-$h$ region, whereas the normalized MVD $\widetilde D_{\rm MV}$ is enhanced predominantly on the small-$h$ side. This contrast indicates that the large-$h$ enhancement of $D_{\rm MV}$ is mainly associated with the overall metric scale rather than with geometric independence. These quantities are also compared with conventional non-equilibrium observables. Although these geometric observables show moderate correlations with conventional observables, such as the half-chain entanglement entropy and the Loschmidt echo, they remain only partially correlated and therefore provide complementary information about non-equilibrium dynamics.

The present framework is not restricted to the transverse-field Ising model and can be extended to more complex quantum many-body systems; see Appendix~\ref{sec:appendix_Potts} as an example.
Possible applications may include strongly interacting QCD matter, such as quark--gluon plasma (QGP); see Refs.~\cite{Busza:2018rrf,Muller:2025qof} as a review of QGP.
The present framework therefore provides a general information-geometric approach to characterizing non-equilibrium quantum-state evolution.

\begin{figure*}[t]
\centering
\includegraphics[width=0.32\linewidth]{figures/heatmap_DMV_L8.pdf}
\includegraphics[width=0.32\linewidth]{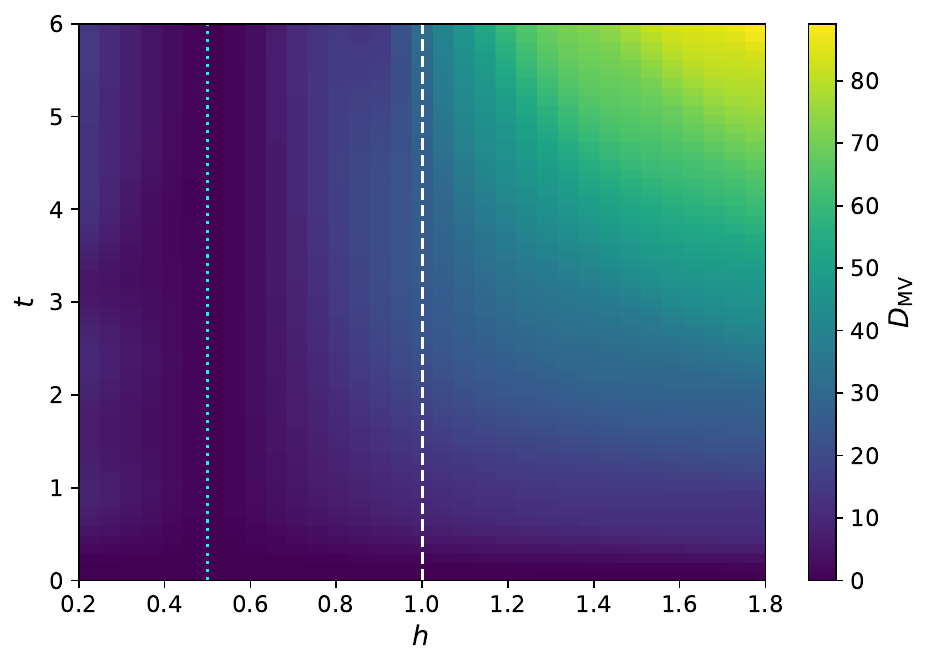}
\includegraphics[width=0.32\linewidth]{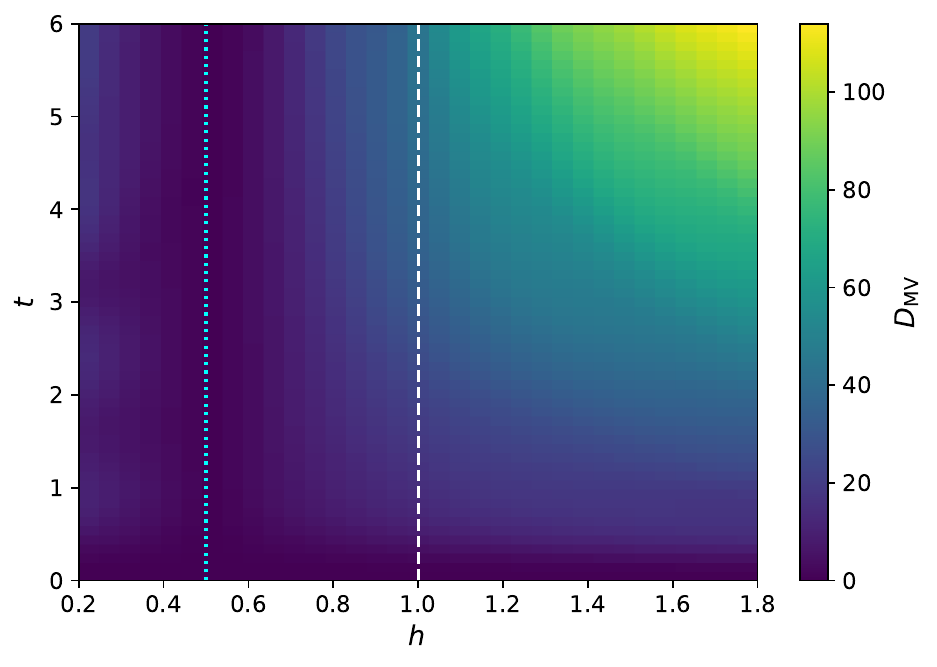}\\
\includegraphics[width=0.32\linewidth]{figures/heatmap_rho_ht_L8.pdf}
\includegraphics[width=0.32\linewidth]{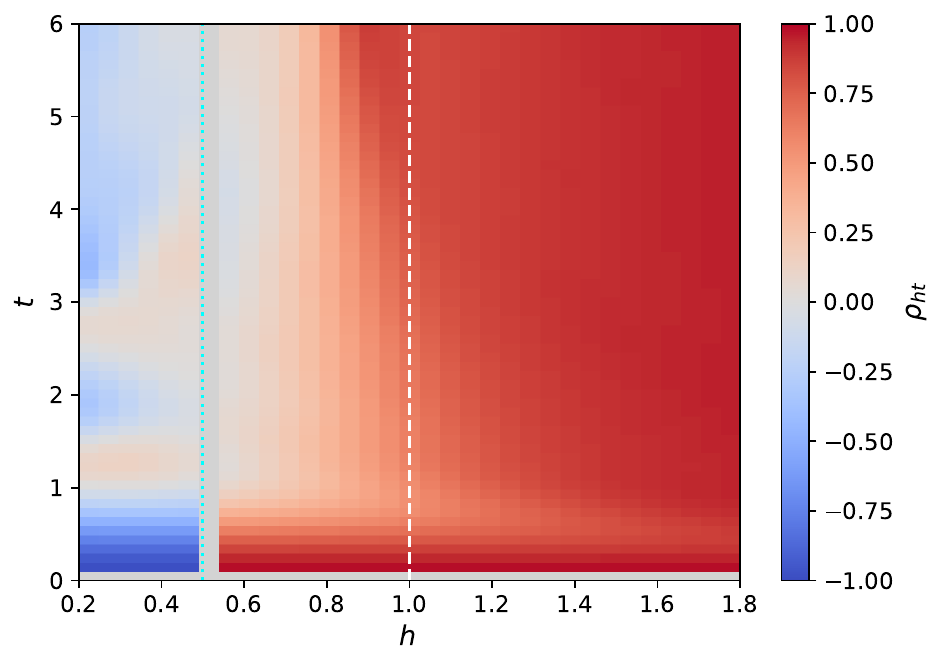}
\includegraphics[width=0.32\linewidth]{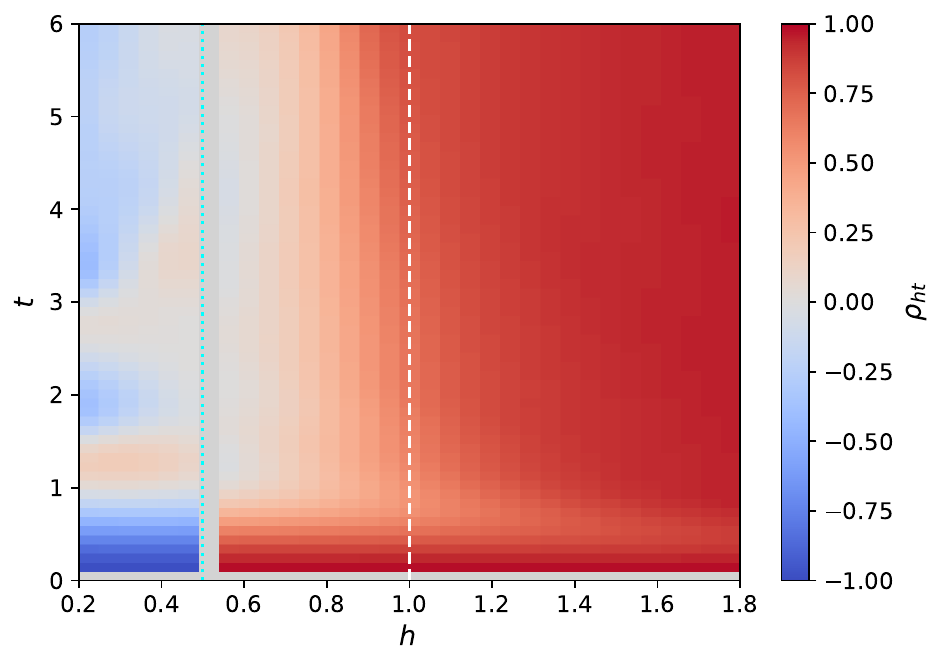}
\caption{
The $L$-dependence of $D_\mathrm{MV}$ (upper panel) and $\rho_{ht}$ (bottom panel) on the information manifold $(h,t)$.
From the left to the right panel, results with $L=8$, $12$ and $16$ are shown, respectively.
The gray region indicates masked areas where the normalization becomes ill-defined because of the no-quench condition.
The dashed and dotted lines are the same as in Fig.~\ref{fig:gmetric}.
}
\label{fig:heatmap_L}
\end{figure*}
\begin{figure}[t]
\centering
\includegraphics[width=1.0\linewidth]{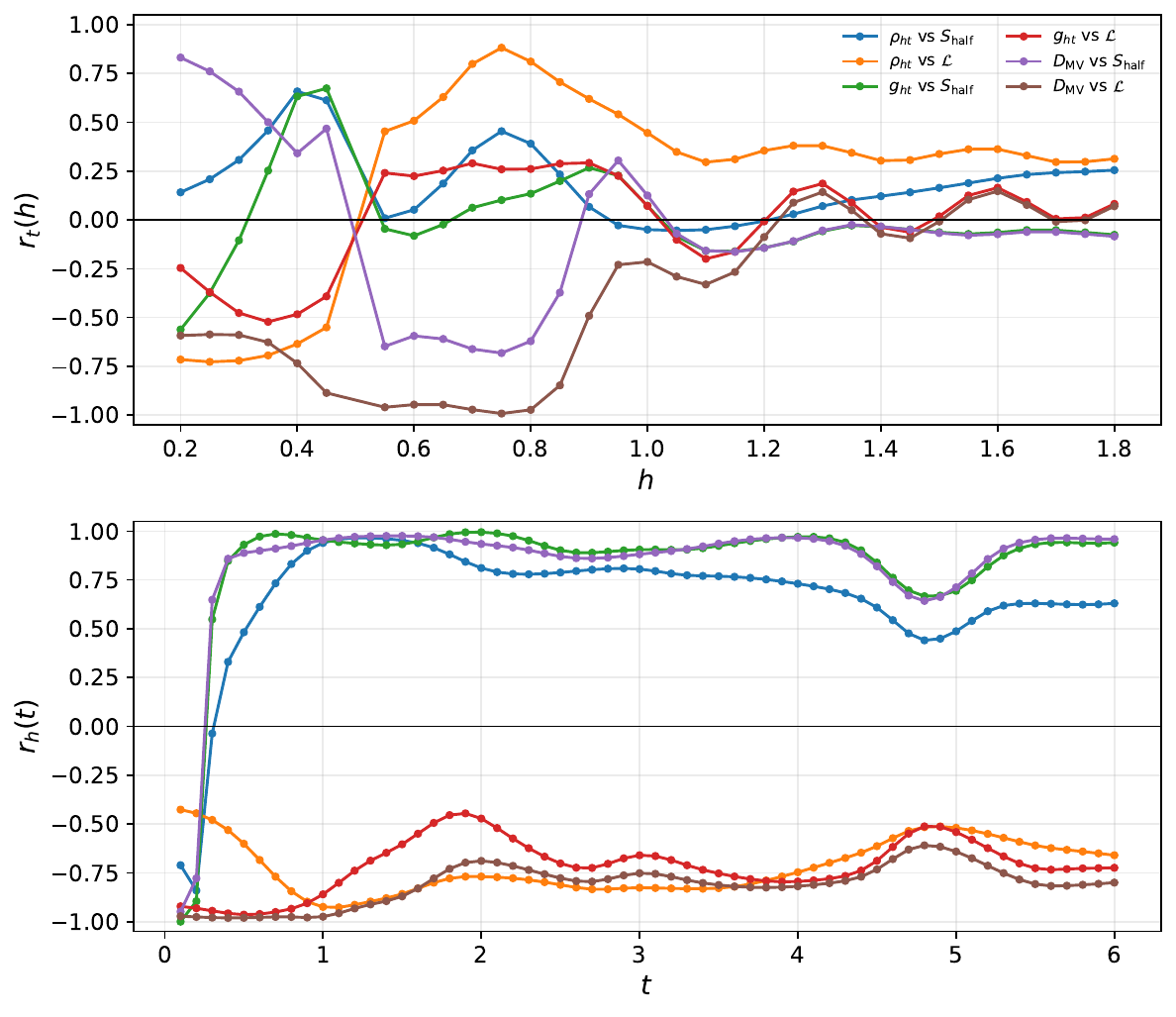}
\caption{
Direction-decomposed Pearson correlation coefficients between geometric observables and conventional observables.
The upper panel shows $r_t(h) = \mathrm{corr}_t[X(h,t),Y(h,t)]$, where the correlation is evaluated over the $t$ direction at fixed $h$.
The lower panel shows $r_h(t)=\mathrm{corr}_h[X(h,t),Y(h,t)]$, where the correlation is evaluated over the $h$ direction at fixed $t$.
}
\label{fig:correlations}
\end{figure}

\begin{acknowledgments}
This work is supported by the Japan Society for the Promotion of Science (JSPS) KAKENHI Grant Numbers JP22H05112 and by the National Natural Science Foundation of China (NSFC) under grant numbers W2433010.
\end{acknowledgments}

\appendix
\section{System-volume dependence}
\label{sec:system_volume}
In this section, to check the dependence of system size on geometric quantities, we show $D_\mathrm{MV}$ and $\rho_{ht}$ on the information manifold $(h,t)$ with $L=8$, $12$ and $16$ as an example.
The results are shown in Fig.~\ref{fig:heatmap_L}.
Relatively strong differences are visible near the critical point $h=1$, which is valid in the thermodynamic limit.
Nevertheless, the overall behavior is qualitatively similar, suggesting convergence toward the thermodynamic limit.
Outside of the critical region, the overall structure of $\rho_{ht}$ remains largely unchanged.
These observations indicate that the geometric features discussed in the main text are robust against finite-size effects.

The present comparison establishes qualitative robustness of the normalized alignment pattern, but does not constitute a finite-size scaling analysis of the metric magnitudes.

\section{Details for Pearson correlation matrix}
\label{sec:correlations}
To clarify whether observed correlations originate from the real-time evolution or from the dependence on the post-quench parameter, we evaluated directional Pearson correlation coefficients.
For a given observable pair $(X,Y)$, we evaluate the coefficient $r_t(h)$ by varying $t$ at fixed $h$, and $r_h(t)$ by varying $h$ at fixed $t$.

Figure~\ref{fig:correlations} shows the Pearson correlation coefficients $r_t$ and $r_h$ as a function of $t$ and $h$.
The correlations along the $t$ direction are generally more variable and may even change sign.
This behavior indicates that the geometric observables and the conventional observables do not generally follow the same real-time evolution, even when they exhibit similar overall trends in the two-dimensional $(h,t)$ parameter space.
In contrast, the correlations evaluated along the $h$ direction are generally much stronger than those obtained along the $t$ direction.
In particular, the mixed metric component $g_{ht}$ and the alignment coefficient $\rho_{ht}$ exhibit very strong positive correlations with the half-chain entanglement entropy $S_\mathrm{half}$ when the correlation is evaluated as a function of $h$ at fixed $t$.
The enhancement of the mixed metric structure is closely associated with the parameter-space dependence of entanglement.
The Loschmidt echo ${\cal L}$ exhibits an opposite behavior.
Both $g_{ht}$ and $\rho_{ht}$ show predominantly negative correlations with ${\cal L}$, particularly along the $h$ direction.
This tendency implies that states which become less similar to the initial state tend to display stronger $t$-$h$ coupling on the information manifold.
The metric-volume density $D_{\mathrm{MV}}$ behaves differently from $g_{ht}$ and $\rho_{ht}$.

\section{Application example}
\label{sec:appendix_Potts}

\begin{figure}[t]
\centering
\includegraphics[width=\linewidth]{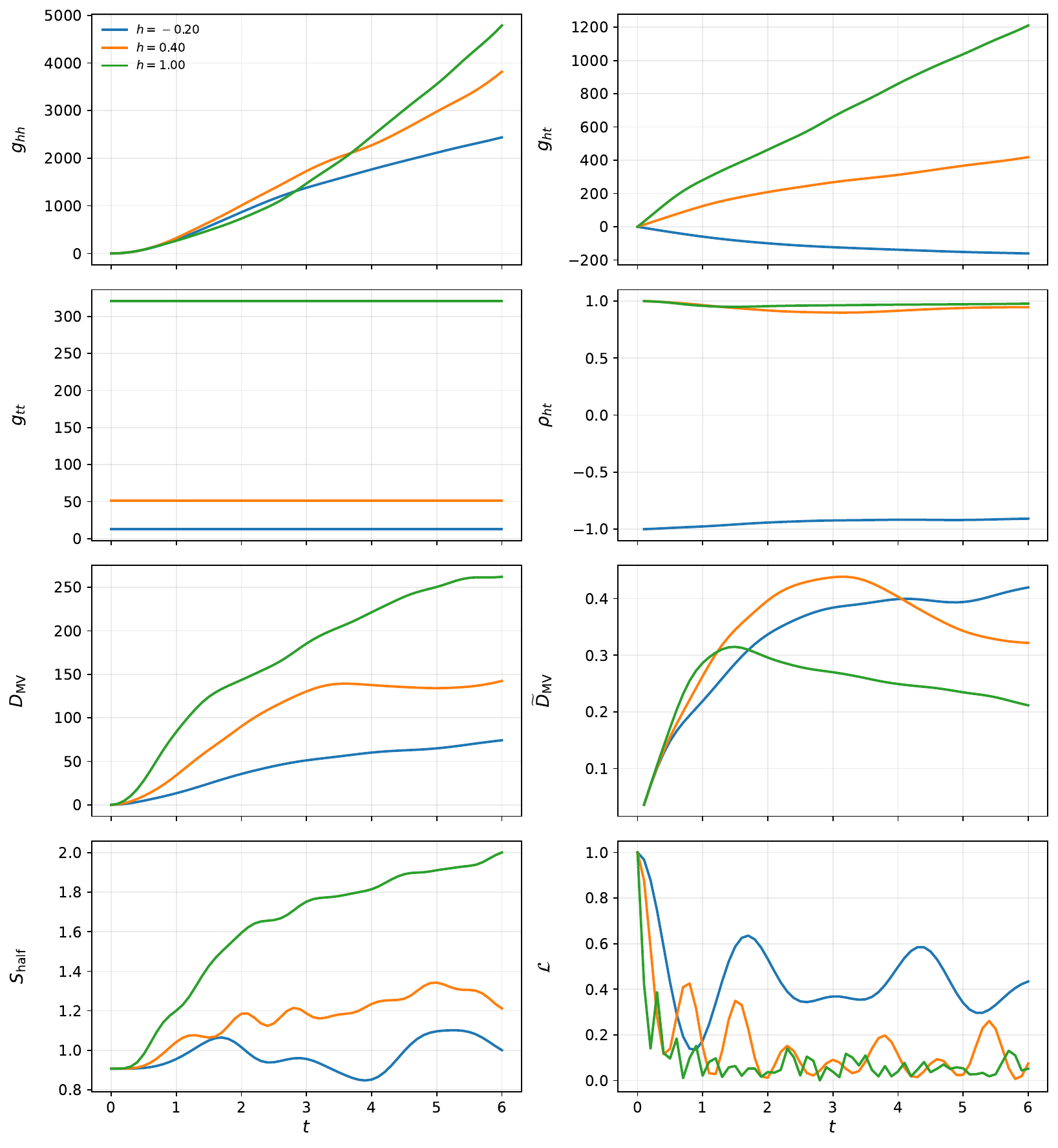}
\caption{
The real-time evolution of  $g_{hh}$, $g_{ht}$, $g_{tt}$, $D_\mathrm{MV}$, $\widetilde{D}_\mathrm{MV}$, $\rho_{ht}$, $S_\mathrm{half}$, and ${\cal L}$ for $h=-0.2$, $0.4$, and $1.0$ in the one-dimensional ${\cal Z}_3$ quantum Potts chain model.
These values of $h$ correspond to quenches into the ferromagnetic region, near the initial value, and into the paramagnetic region, respectively.
}
\label{fig:timecuts_Potts}
\end{figure}

To demonstrate that the geometric diagnostics are not restricted to the transverse-field Ising model, we briefly apply the same framework to the one-dimensional $\mathbb{Z}_3$ quantum Potts chain model and also the lattice Schwinger model.
Note that the $\mathbb{Z}_3$ quantum Potts chain model has some relations with the QCD-like Potts models considered in Refs.~\cite{Alford:2001ug,Kim:2005ck,Kashiwa:2020waa}.

\subsection{$\mathbb{Z}_3$ quantum Potts model}
The Hamiltonian of the $\mathbb Z_3$ quantum Potts model with the open boundary condition is defined as
\begin{align}
    H=
    & - J \sum_{i=1}^{L-1} \left( Z_i Z_{i+1}^{\dagger} + Z_i^{\dagger} Z_{i+1} \right)
    \nonumber\\
    & - \Gamma \sum_{i=1}^{L} \left( X_i+X_i^{\dagger} \right)
      - h \sum_{i=1}^{L} \left( Z_i + Z_i^{\dagger} \right),
\end{align}
where $Z_i$ and $X_i$ are the local $\mathbb{Z}_3$ operators satisfying
\begin{align}
    Z_i^3 &= X_i^3 = 1,
    ~~~
    X_i Z_i = \omega Z_i X_i,
\end{align}
with $\omega = e^{2\pi i/3}$.
Here, we set $L=8$, $J=1$, and $\Gamma=0.8$.
The term proportional to $\Gamma$ is introduced to mimic quantum fluctuations analogous to those expected in QCD.
 Currently, the value of $\Gamma$ is treated as a free parameter.
The parameter $h$ acts as an external field that explicitly breaks the $\mathbb{Z}_3$ symmetry.

The initial state is prepared as the ground state of $H(h_i)$;
\begin{align}
    | \psi_{\mathrm{GS}}(h_i) \rangle.
\end{align}
Following the same procedure as in the main text, we construct the time-evolved state as
\begin{align}
    |\psi(h,t)\rangle &= e^{-iH(h)t} |\psi_\mathrm{GS}(h_i) \rangle ,
\end{align}
and evaluate the QFI metric on the extended information manifold $(h,t)$.
Here, we set the initial value to $h_i=0$.
Note that the external field $h$ consists of the current quark mass $m_0$ and the quark chemical potential $\mu$~\cite{Alford:2001ug,Kim:2005ck,Kashiwa:2020waa}.
Consequently, we may consider information manifolds such as $(m_0,t)$ and $(\mu,t)$ in QCD and related effective models.

Figure~\ref{fig:timecuts_Potts} shows the real-time evolution of the geometric observables for several values of $h$.
As $h$ increases, $g_{hh}$ exhibits a pronounced enhancement, indicating that the evolved state becomes sensitive to variations of the post-quench parameter.
At the same time, $g_{ht}$ grows significantly, demonstrating that parameter deformation and real-time evolution are no longer geometrically independent.
This tendency is further reflected in $|\rho_{ht}|$, whose magnitude approaches unity over broad time intervals.
Consequently, the $h$ and $t$ directions become strongly aligned on the information manifold. Consistent with this enhanced alignment, $\widetilde D_{\rm MV}$ decreases, indicating a reduction of the effective geometric independence between the parameter deformation and the real-time evolution.

\begin{figure}[t]
\centering
\includegraphics[width=\linewidth]{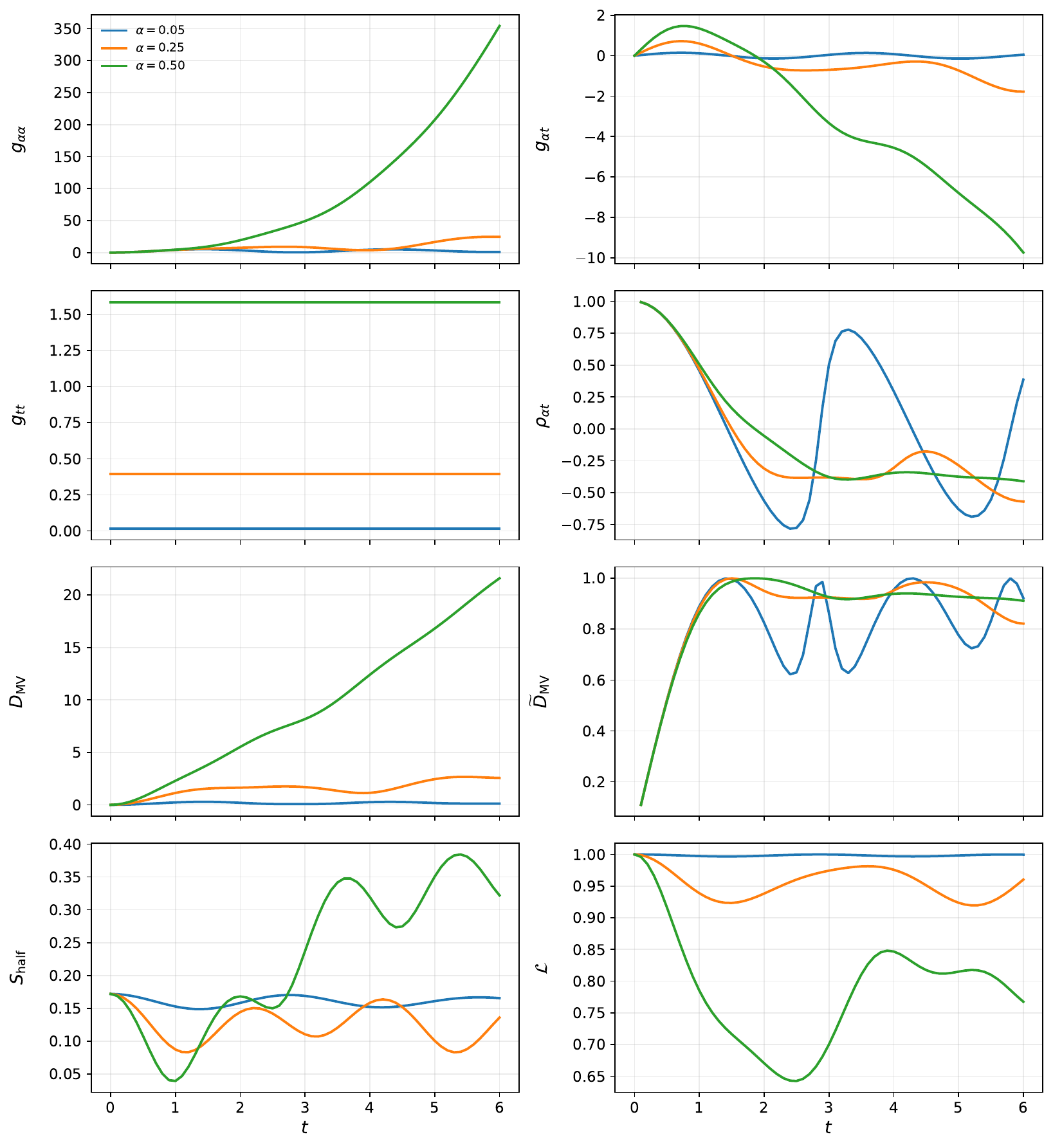}
\caption{
The real-time evolution of $g_{\alpha \alpha}$, $g_{\alpha t}$, $g_{tt}$, $D_\mathrm{MV}$, $\widetilde{D}_\mathrm{MV}$, $\rho_{\alpha t}$, $S_\mathrm{half}$, and ${\cal L}$ for $\alpha=0.05$, $0.25$, and $0.5$ in the one-dimensional Schwinger model.
}
\label{fig:timecuts_Schwinger}
\end{figure}

\subsection{Schwinger model}
As an additional example, we consider the lattice Schwinger model~\cite{schwinger1962gauge,Kogut:1974ag,Banuls:2019bmf} with the open boundary condition. After eliminating the gauge degrees of freedom through Gauss's law, the Hamiltonian is written as
\begin{align}
    H(\alpha)
    & = \frac{x}{2} \sum_{n=0}^{L-2} \left( \sigma_n^x\sigma_{n+1}^x + \sigma_n^y\sigma_{n+1}^y \right)
    \nonumber\\
    & + \frac{m}{2} \sum_{n=0}^{L-1} (-1)^n\sigma_n^z
      + \sum_{n=0}^{L-2} \left( \alpha+\sum_{k=0}^{n}q_k \right)^2,
\end{align}
where the staggered fermion charge density is
\begin{align}
    q_k &= \frac{1}{2} \Bigl[ \sigma_k^z + (-1)^k \Bigr],
\end{align}
and $L$ means the system size, $x$ denotes the hopping parameter, $m$ is the staggered
fermion mass, and $\alpha$ represents the background external electric-field.
In this study, $\alpha$ is treated as the controllable Hamiltonian parameter in the model.
Here, we set $L=10$, $x = 1$ and $m = 0.5$.
Note that $\alpha$ is periodic, $\alpha \equiv \alpha + 1$.

The initial state is prepared as the ground state of $H(\alpha_i)$;
\begin{align}
    |\psi_{\mathrm{GS}}(\alpha_i)\rangle.
\end{align}
The non-equilibrium dynamics is generated by the post-quench Hamiltonian with external field $\alpha$ as
\begin{align}
    |\psi(\alpha,t)\rangle &= e^{-iH(\alpha)t} |\psi_{\mathrm{GS}}(\alpha_i)\rangle.
\end{align}
and evaluate the QFI metric on the extended information manifold $(\alpha,t)$.
Here, we set the initial value to $\alpha_i=0$.

Figure~\ref{fig:timecuts_Schwinger} shows the real-time evolution of geometric observables for several values of $\alpha$.
The diagonal metric component $g_{\alpha\alpha}$ exhibits a pronounced growth as $\alpha$ increases.
While the response remains relatively moderate for $\alpha=0.05$, a substantial enhancement is observed for $\alpha=0.5$, indicating that the quantum state becomes increasingly sensitive to variations in $\alpha$.
A similar tendency is reflected in the metric-volume density $D_{\rm MV}$, whose magnitude increases significantly with increasing $\alpha$.
These results suggest that stronger $\alpha$ amplify the information-geometric response of the dynamics.
The alignment coefficient $\rho_{\alpha t}$ develops increasingly large absolute values at larger $\alpha$.
This behavior indicates an enhanced geometric correlation between the $\alpha$ direction and the $t$ direction on the information manifold.
Consistently, the normalized MVD decreases as the alignment becomes stronger, reflecting a reduction of the effective geometric independence of the tangent directions.

\bibliography{ref.bib}

\end{document}